\pdfoutput=1
\documentclass[journal,twoside,web]{ieeecolor}
\usepackage[pdftex]{graphicx}
\usepackage[outdir=./]{epstopdf}
\usepackage{generic}
\usepackage{cite}
\usepackage{amsmath,amssymb,amsfonts}
\usepackage{algorithmic}
\usepackage{graphicx}
\usepackage{textcomp}
\usepackage{systeme}
\usepackage{float}
\usepackage{verbatim}
\usepackage{dblfloatfix} 
\def\BibTeX{{\rm B\kern-.05em{\sc i\kern-.025em b}\kern-.08em
    T\kern-.1667em\lower.7ex\hbox{E}\kern-.125emX}}
\begin{document}
\title{Energy-Efficient Implementation of Generative Adversarial Networks on Passive RRAM Crossbar Arrays}
\author{Siddharth Satyam, Honey Nikam and Shubham Sahay, \IEEEmembership{Member, IEEE}
\thanks{Siddharth Satyam and Honey Nikam are with the Department of Mechanical Engineering, Indian Institute of Technology Kanpur, Kanpur 208016, India}
\thanks{Shubham Sahay is with the Department of Electrical Engineering, Indian Institute of Technology Kanpur, Kanpur 208016, India (e-mail: ssahay@iitk.ac.in). This
        work was partially supported by the Semiconductor Research Corporation
        (SRP Task 3056.001) and IIT Kanpur.}}
\maketitle

\begin{abstract}
Generative algorithms such as GANs are at the cusp of next revolution in the field of unsupervised learning and large-scale artificial data generation. However, the adversarial (competitive) co-training of the discriminative and generative networks in GAN makes them computationally intensive and hinders their deployment on the resource-constrained IoT edge devices. Moreover, the frequent data transfer between the discriminative and generative networks during training significantly degrades the efficacy of the von-Neumann GAN accelerators such as those based on GPU and FPGA. Therefore, there is an urgent need for development of ultra-compact and energy-efficient hardware accelerators for GANs. To this end, in this work, we propose to exploit the passive RRAM crossbar arrays for performing key operations of a fully-connected GAN: (a) true random noise generation for the generator network, (b) vector-by-matrix-multiplication with unprecedented energy-efficiency during the forward pass and backward propagation and (C) in-situ adversarial training using a hardware friendly Manhattan's rule. Our extensive analysis utilizing an experimentally calibrated phenomological model for passive RRAM crossbar array reveals an unforeseen trade-off between the accuracy and the energy dissipated while training the GAN network with different noise inputs to the generator. Furthermore, our results indicate that the spatial and temporal variations and true random noise, which are otherwise undesirable for memory application, boost the energy-efficiency of the GAN implementation on passive RRAM crossbar arrays without degrading its accuracy.
\end{abstract}

\begin{IEEEkeywords}
Generative Adversarial Networks, Passive RRAM crossbar, Manhattan's rule, True random noise.
\end{IEEEkeywords}

\section{Introduction}
The unprecedented development in the field of supervised deep learning models, which rely on large labelled or annotated data sets, has transformed almost all the facets of human endeavor in this era of internet of things (IoT) and big data. However, their application is limited in domains where generating such a labelled data is extremely difficult or costly. Therefore, unsupervised and semi-supervised generative models, which may learn the patterns or features in a complicated data set and generate high-quality artificial (fake) data bearing similar features as that of the original data set, have been extensively explored \cite{b1,b2}. Generative Adversarial Networks (GANs), a subclass of generative models, are considered one of the most promising approaches towards large-scale synthetic data generation and unsupervised/semi-supervised learning \cite{b1,b2}. GANs have been applied to solve a wide variety of problems including image synthesis and super-resolution, 3D object generation, image-to-text translation (and vice-versa), image-to-image translation, speech recognition, attention prediction, autonomous driving, etc. \cite{b1,b2,b3}. In GANs, two adversarial (competitive) networks are co-trained alternately: the generator network is trained to produce artificial (fake) data which may not be distinguishable from the original data set; while the discriminator network is trained to classify whether the input data belongs to the original data set or obtained from the generator network. The training methodology for GAN involves movement of large amount of data between the two adversarial networks leading to a significant memory and computational resource consumption \cite{b3,b4,b5,b6,b7,b8,b9} which restricts their application in IoT edge devices with limited energy and area. Therefore, development of a compact and ultra-low power GAN processing engine is indispensable for enabling unsupervised learning and generating artificial data on resource-constrained IoT edge devices.

The digital GAN accelerators based on FPGA and GPUs exhibit significantly high latency and energy consumption owing to the intensive data shuffling between the storage and computational blocks due to their inherent von-Neumann architecture. Since vector-by-matrix multiplication (VMM) is the fundamental operation in GANs during the forward pass and backward propagation using optimization algorithms such as gradient descent, RMSprop, ADAM, etc. \cite{b1,b2,b3}, the data shuffling and the energy consumption while training GANs can be significantly reduced by performing in-memory VMM operations exploiting cross-point arrays of emerging non-volatile memories. Recently, several innovative deep convolutional (DC) GAN architectures including layer-wise pipelined computations \cite{b4}, efficient deconvolutional operation \cite{b5}, computational deformation technique to facilitate efficient utilization of computational resources in transpose convolution \cite{b6}, and a ternary GAN \cite{b7} utilizing in-memory VMM engines based on RRAMs (with binary and 2-bit storage capability) and SOT-MRAMs were proposed. Moreover, a hybrid CMOS-analog RRAM-based implementation of DCGAN (without the pooling layer) including digital error propagation and weight update units was also proposed \cite{b8}. However, the non-linearity in RRAM conductance update during the training process was not considered, and the weight sign crossbar and sequential reading of output limits the efficacy of the analog DCGAN implementation.

Recently, a fully-connected GAN implementation utilizing the intrinsic read noise (conductance fluctuation during the read operation) and write noise (imprecise conductance tuning during the write operation) of active (1T-1R) RRAM crossbar array was experimentally demonstrated in \cite{b9} for generating (3 classes of) artificial digital patterns of handwritten digits after training on reduced MNIST dataset. An optimal level of write noise was found to increase the diversity in the generated patterns and mitigate the mode dropping issue \cite{b9}. Although the selector MOSFET in the active (1T-1R) RRAM configuration reduces the cell leakage current, improves the tuning precision, provides current compliance, and facilitates partial/selective programming of the array for large-scale hardware demonstrations of neuromorphic networks \cite{b10,b11,b12,b13,b14}, it also leads to a significantly large area overhead. The scalability of the selector MOSFET is limited since it has to provide large programming/forming currents to the RRAM device and sustain high voltages during forming/write operation\cite{b15}.  

On the other hand, the passive RRAM crossbar arrays exhibit a significantly reduced area, lower fabrication complexity and cost, and an inherent scaling benefit since they do not require a selector MOSFET \cite{b15,b16,b17,b18,b19,b20,b21}. However, unlike active (1T-1R) RRAM cells, the passive RRAM crossbar cells are susceptible to sneak path leakage currents and half-select cell disturbance \cite{b16,b17}. Moreover, the spatial variation in the switching threshold and limited crossbar yield degrades their performance \cite{b16,b17}. Nevertheless, the recent advancements in the fabrication process and material stacks for RRAMs, novel programming schemes and conductance mapping techniques have enabled the realization and CMOS BEOL integration of large passive RRAM crossbar arrays with high conductance tuning precision, large yield and uniformity, highly non-linear characteristics and significantly suppressed sneak path leakage current \cite{b15,b18,b19,b20}. Considering the promising scaling prospects of the passive RRAM crossbar arrays, it becomes imperative to explore their potential for implementation of GANs. Moreover, the inherent spatial and temporal variations of the passive RRAM crossbar array can be explored to extract a true random noise source for the generator network.

To this end, in this work, we develop a hardware-aware simulation framework and demonstrate a compact and ultra-energy efficient fully-connected GAN utilizing passive RRAM crossbar arrays for synthetic image generation. We propose a methodology to generate true random noise using passive RRAM crossbar array for input to the generator network and perform the VMM during the forward pass and backward propagation, and in-situ adversarial training of GAN using a hardware friendly Manhattan's rule (fixed pulse training) on passive RRAM crossbar arrays. While most hardware solutions aimed towards efficient implementation of GANs focus on discriminative networks, we also investigated the impact of different (pseudo random and true random) noise input to the generator network on the training energy and accuracy of the fully-connected GAN. Our extensive analysis utilizing an experimentally calibrated phenomological model for passive RRAM crossbar array (which accurately captures the non-ideal effects such as spatial and temporal device variations and noise) indicates that the proposed GAN implementation with true random noise input to the generator and device-to-device variations in the passive RRAM crossbar array exhibits a significantly enhanced energy-efficiency with accuracy comparable to the software implementation. Our results may provide incentive for experimental demonstration of GANs on passive RRAM crossbar arrays.  

The manuscript is organized as follows: Sections I.A and I.B provide a brief overview of the fully-connected GANs and passive RRAM crossbar arrays, respectively. The intricate details regarding the simulation framework developed in this work utilizing an experimentally calibrated phenomological model for passive RRAM crossbar array are discussed in section II. The performance estimates of the proposed GAN implementation using passive RRAM crossbar arrays for important metrics such as accuracy, diversity in the generated images, area and energy are reported in section III while the conclusions are drawn in section IV.

\begin{figure}[b]
    \centering
    \includegraphics[width=0.46\textwidth]{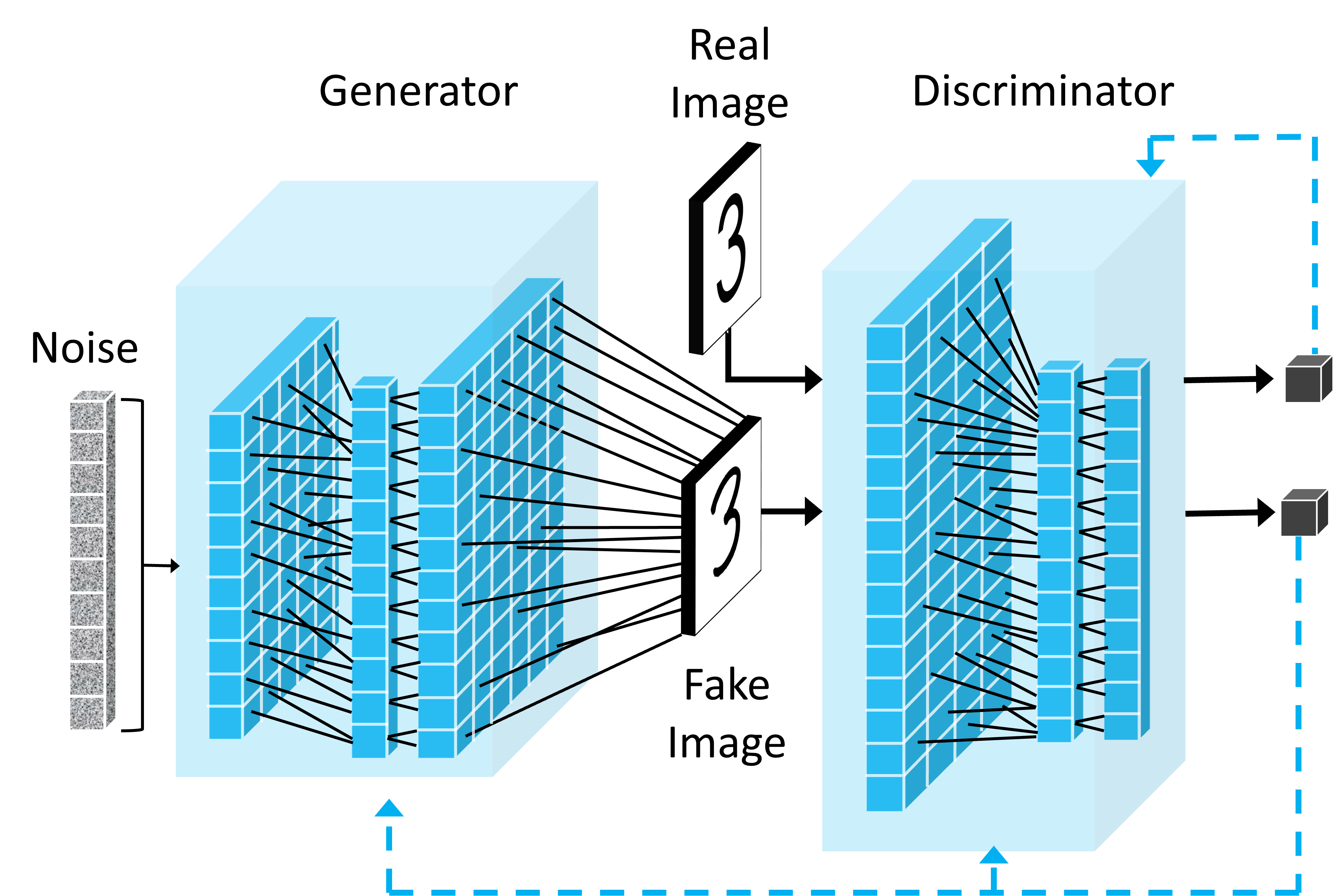}
    \caption{The schematic representation of a fully-connected GAN. The generator produces a fake image while the discriminator gives the output probability for both real and fake image inputs (represented by the black boxes). These are fed back during backward propagation (represented by the dashed blue lines).}
\end{figure}

\subsection{Generative Adversarial Networks (GANs)}
Fully-connected GANs rely on an adversarial training process that involves a zero-sum game between two multi-layer perceptrons: the generator and the discriminator network as shown in Fig. 1. Two conflicting objectives must be fulfilled while training a fully-connected GAN: the discriminator should be able to predict accurately whether the data produced by the generator is real or fake, and the generator should be able to synthesize artificial data that contains features which are indistinguishable from the original data and deceive the discriminator to classify it as real data. The generator network is fed with a noise input and produces a fake image based on its weights and parameters. These fake images are fed to the discriminator along with the real images and the discriminator assigns appropriate labels to these images (whether fake or real). The cost function used for training can be formulated as \cite{b2}:
\begin{align}
   Cost = \mathbb{E}_{x \sim P_{data}}log(D(x)) +  \mathbb{E}_{z \sim P_{z}}log(1-D(G(z)))
\end{align}

where $G(z;W_g$) represents the mapping of an input vector consisting of a randomly distributed data from the noise variable $P_z$ to the generator output based on the parameters $W_g$ and $D(x;W_d)$ represents a mapping from the data space to a scalar quantity which indicates the probability that the input image to the discriminator belongs to the real data set (or the artificial data provided by the generator). The cost function represented by equation (1) is a "minimax" game in which the generator tries to minimise the cost (such that $D(G(z)) \rightarrow 1$ and $Cost \rightarrow -\infty$) while the discriminator tries to maximise the cost ($D(G(z)) \rightarrow 0$) and the game concludes when the system reaches the Nash equilibrium \cite{b9}.

\begin{figure}[t]
    \centering
    \includegraphics[width=\linewidth]{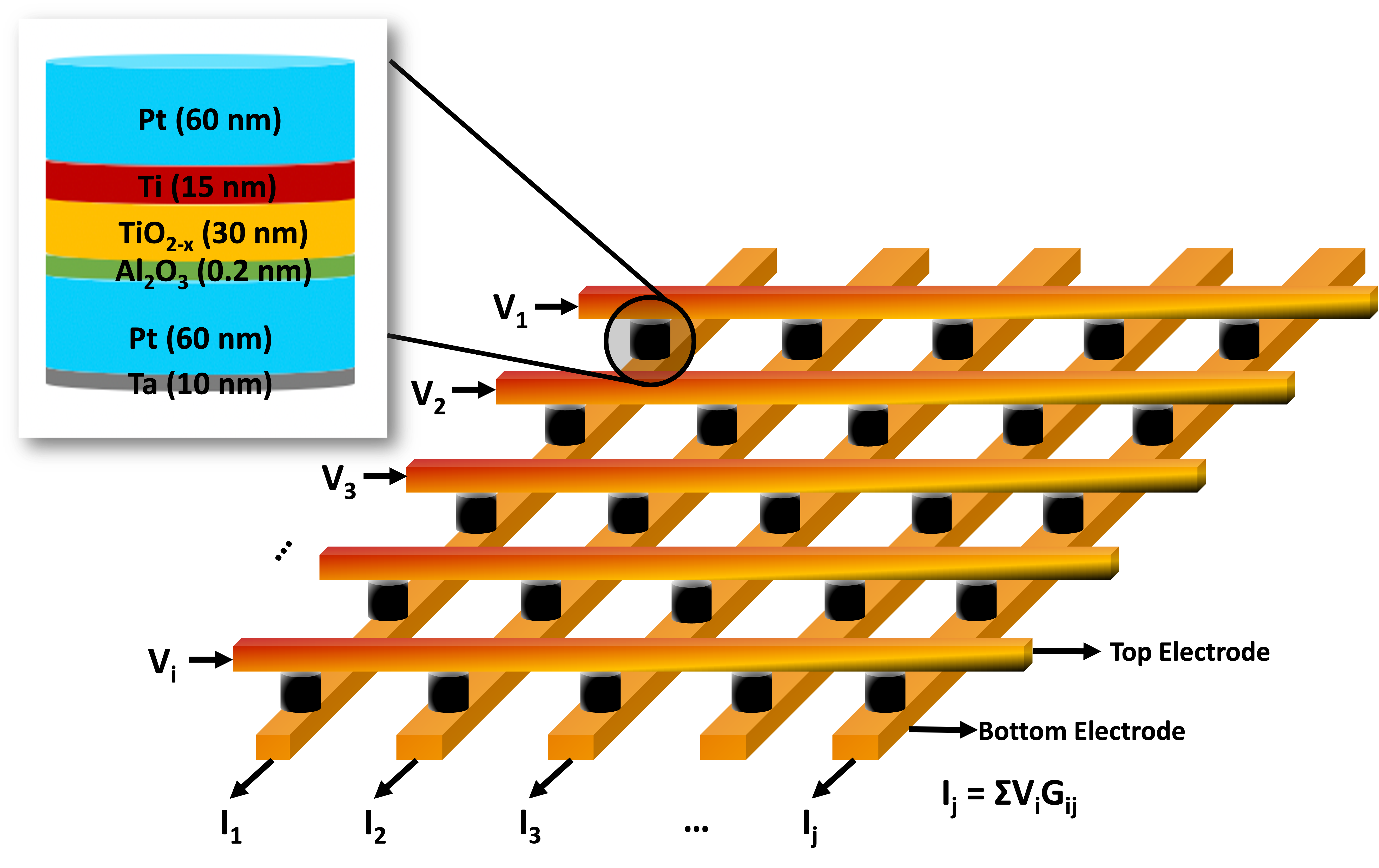}
    \vspace{-20pt}
    \caption{3D view of the passive RRAM crossbar array based on Pt/Al$_2$O$_3$/TiO$_{2-x}$/Ti/Pt stack used in this work.}
\vspace{-10pt}
\end{figure}

\subsection{Passive RRAM crossbar arrays}
Filamentary resistive RAMs (RRAMs) are metal-insulator-metal (MIM) structures in which the insulator (typically a non-stoichiometric transition metal oxide) exhibits a reversible switching between different resistance-states with the aid of electrical pulses. RRAM devices can be arranged in two configurations: (a) active 1T-1R crossbar where the RRAM devices are integrated on top of the drain of a selector MOSFET which provides efficient current compliance and precise conductance-tuning capability at the cost of an increased area overhead, and (b) passive crossbar array in which the RRAMs are realized at the intersection of orthogonal word lines (WLs) and bit lines (BLs) as shown in Fig. 2. The passive RRAM crossbar provides an inherent scaling benefit since the footprint of the RRAM device is dictated by the width of the metal interconnects rather than a selector MOSFET. Moreover, the VMM operation can be performed in-situ on a passive RRAM crossbar array exploiting physical laws with an unprecedented energy-efficiency by encoding inputs as WL voltages and the weights as the conductance of the RRAM devices \cite{b22,b23,b24,b25}. 

\section{Modeling Approach and Simulation Framework}

Considering the scaling prospects and ultra-high energy-efficiency of the passive RRAM crossbar arrays while performing in-situ computations, we implemented a vanilla GAN to synthesize handwritten digits from the MNIST data set. The MNIST data set is a collection of 28$\times$28 pixel images of handwritten digits which were flattened into 784-dimensional vectors and then normalized to the range [-1,1]. 

The generator network in a GAN model creates a mapping between an input latent space and an output sample space. The inputs are provided from a random distribution (noise) to ensure that the generator takes a different input at each iteration and generates a different instance of output data. The weights of the generator network are then trained considering the output probability of the discriminator corresponding to the input fed by the generator at each iteration. In our implementation, the random input is a 1D-array, and the output is a flattened 2D-image ($28 \times 28$ pixels) arranged as a 1D-array with 784 elements. While most of the prior hardware GAN implementations have focused only on the discriminator network or the deconvolution operation of the generator network, in this work, we have also explored the impact of the input noise distribution of the generator network on the energy consumption and the accuracy of the GAN implementation. We have considered two types of input noise distribution: (a) pseudo-random noise input where the samples are generated using the software pseudo random number generator (PRNG) from a standard normal distribution $\mathcal{N}(0,1)$ with a predefined seed and (b) true-random noise input where the samples are generated exploiting the inherent spatial and temporal variations in the passive RRAM crossbar array.

\begin{figure}[t]
    \centering
    \includegraphics[width=0.5\textwidth]{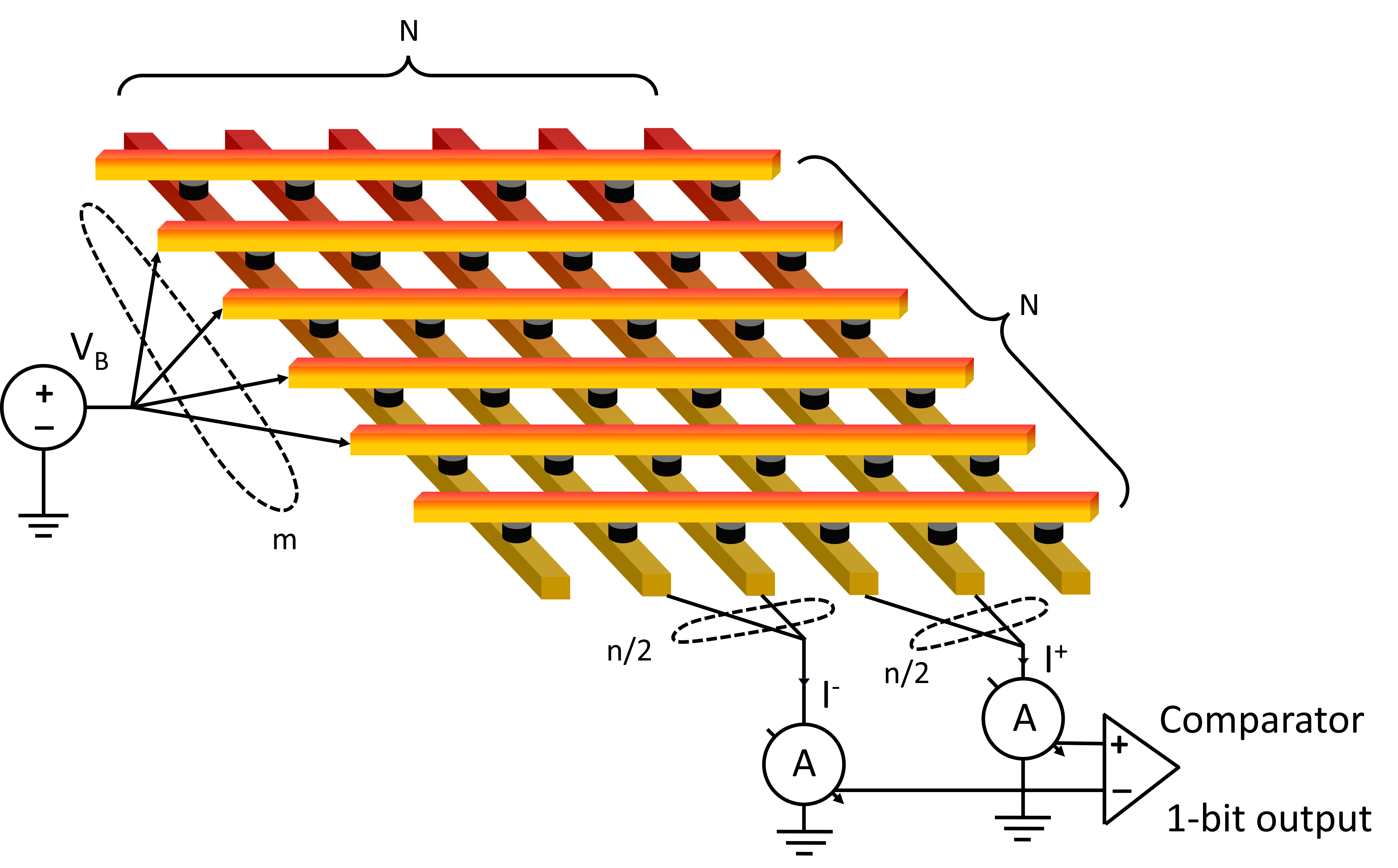}
    \caption{Proposed scheme for generation of true-random noise for the generator network. }
\end{figure}

We propose a novel methodology to generate the random noise inputs from passive RRAM crossbar array as shown in Fig. 3. We randomly select $n$ (out of $N$) columns from the passive RRAM crossbar array in each iteration and divide them into two groups of $n/2$ columns each. We program all the RRAM cells in the two sets of $n/2$ columns using identical write pulses in an attempt to realize cells with same conductance-states (within the range of $150\mu S$ to $300\mu S$) in both groups. However, owing to the spatial (device-to-device) variation in the switching threshold voltage of RRAMs in the array, the RRAM cells at different location in the array exhibit different conductance-states \cite{b16,b26}. The random input bit is then generated by applying a read pulse to $m$ rows (selected randomly out of $N$ rows in each iteration) and comparing the integrated current from the two groups of $n/2$ columns. The temporal read current fluctuation and random telegraph noise (RTN) further add to the entropy and enhance the randomness \cite{b27,b28}. 

We initialise a generator network of dimension [100,128,784] with random weights and feed it with a noise input of size 100 bits generated either through the software PRNG or the proposed true random noise generator (TRNG) utilising the passive RRAM crossbar array. The forward propagation of inputs, represented by the affine transformation $z$ in equation (2), is followed by the application of an activation function $a = f(z)$ at each layer:
\begin{align}
    z = W^T \cdot x + b
\end{align}
where $b$ represents the biases and $W$ represents the weights which are stored in the passive RRAM crossbar array in the form of conductance-states in the proposed GAN implementation. For any practical application, the fully-connected GAN implementations require a large number of weights which cannot be accommodated on a single passive RRAM crossbar array. Therefore, we utilize 54 ($64 \times 64$ \cite{b15}) RRAM crossbar arrays in the proposed GAN implementation. Moreover, for extracting optimal performance, the conductance values of RRAMs are chosen in the range of $G_{min} = 150\mu S$ to $G_{max} = 300\mu S$. Since conductance values are always positive while the weights of a neural network can be bipolar in the software, we propose a novel conductance-to-weight mapping scheme $f:G\to W$ as: 

\begin{align}
W_{ij} = \pm \{ W_{min} + \frac{G_{ij} - G_{min}} {G_{max} - G_{min}} \cdot (W_{max} - W_{min}) \}
\end{align}

where $W_{max}$ and $W_{min}$ are the maximum and minimum weights used during the training process. While $W_{min}$ is kept 0 for both discriminator and generator, we have clipped the $W_{max}$ to 0.4 for the generator and 0.15 for the discriminator. The proposed mapping scheme requires that the weights do not change their sign during the training process i.e. positive weights remain positive and negative weights remain negative throughout the training of GAN. The $+$ sign and $-$ sign from equation (3) represent the mapping of positive and negative weights from the conductances. Using this mapping scheme, we assign separate RRAM cells for encoding the positive and negative weights on the same passive RRAM crossbar unlike the differential scheme where the positive and negative weights are stored in different crossbars.

A 784-dimensional vector which represents the fake image is produced at the output of the generator after the forward propagation. The generator output along with the real image (from flattened 784-dimensional MNIST training data) are then fed to the discriminator network with a dimension of [784,128,1]. The discriminator network generates a scalar output which indicates the probability of the input image being real or fake. The outputs of the discriminator and the generator networks are then used as inputs for the back propagation step to determine the gradients and train the GAN.

\begin{figure}[t]
    \centering
    \includegraphics[width=0.45\textwidth]{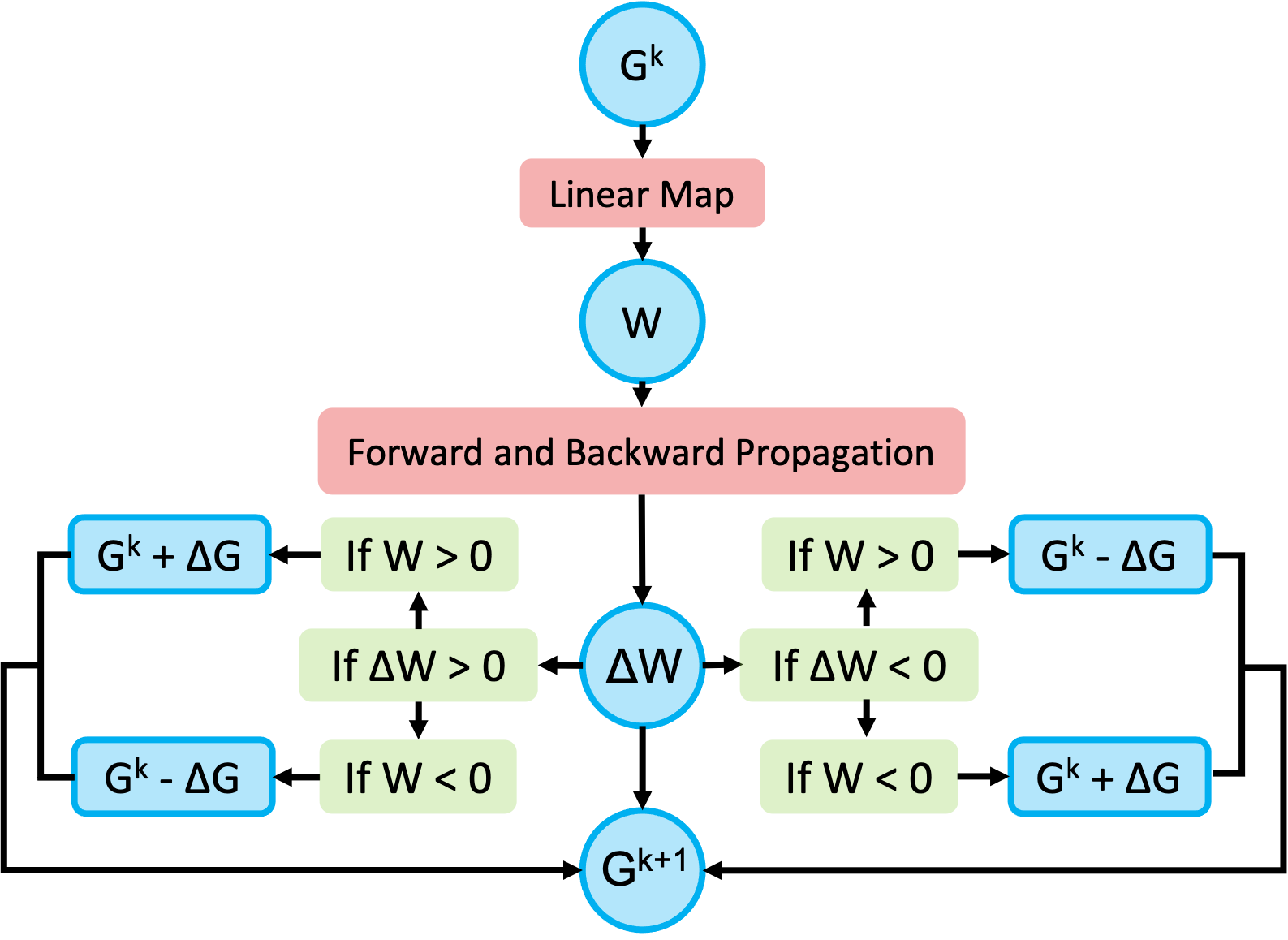}
    \caption{The flowchart of the weight update process used in this work. The superscripts k and k+1 indicate the k$^{th}$ and (k+1)$^{th}$ iterations while $\Delta G$ represents the absolute conductance change.}
\end{figure}

\begin{figure*}[t]
    \centering
    \includegraphics[width=\textwidth]{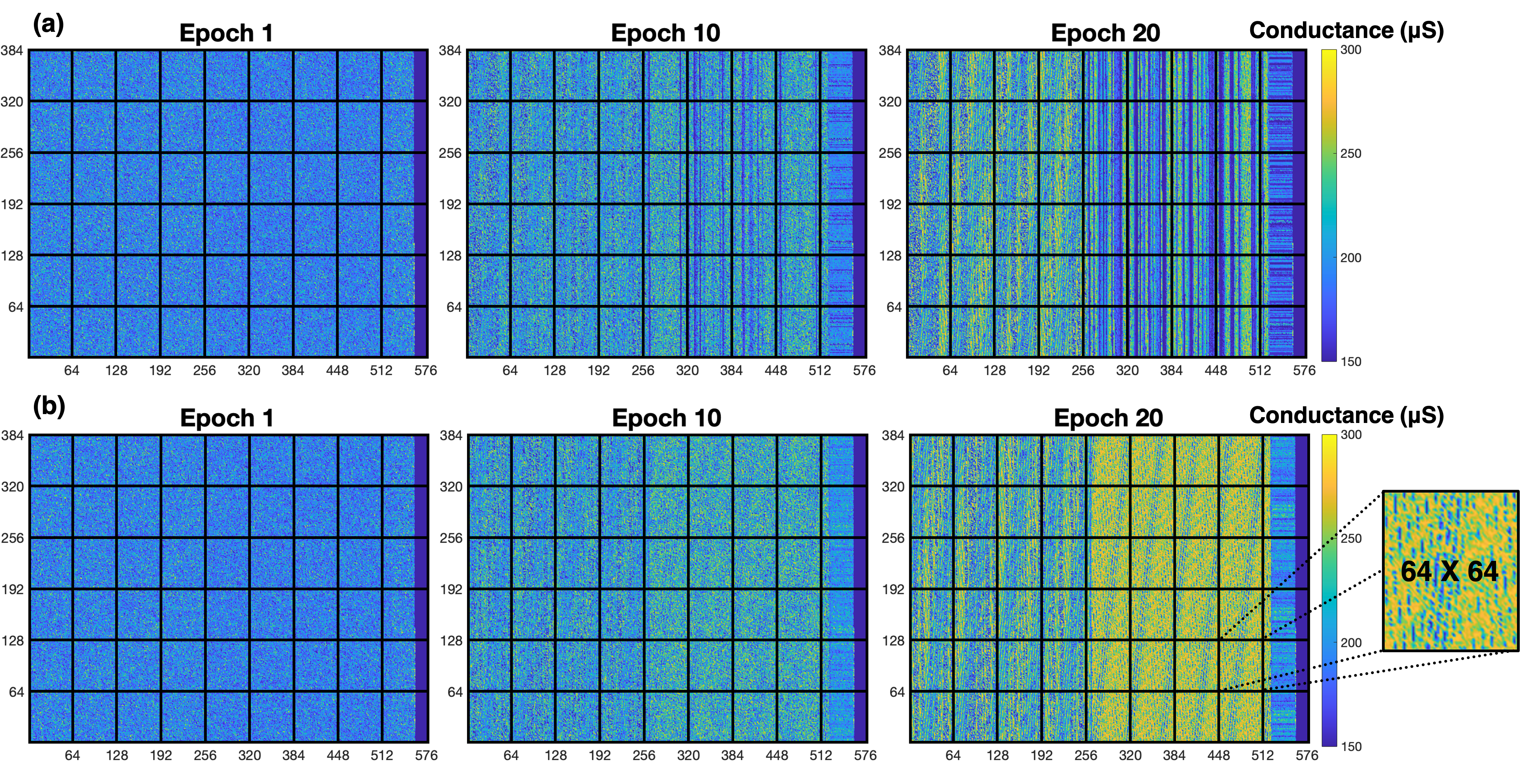}
    \caption{Conductance evolution while training the GAN implemented on 54 ($64\times 64$) RRAM crossbar arrays for generator network with (a) true random noise input considering device-to-device variations and (b) pseudo random noise input.}
\end{figure*}

\begin{figure*}[b]
    \centering
    \includegraphics[width=\textwidth]{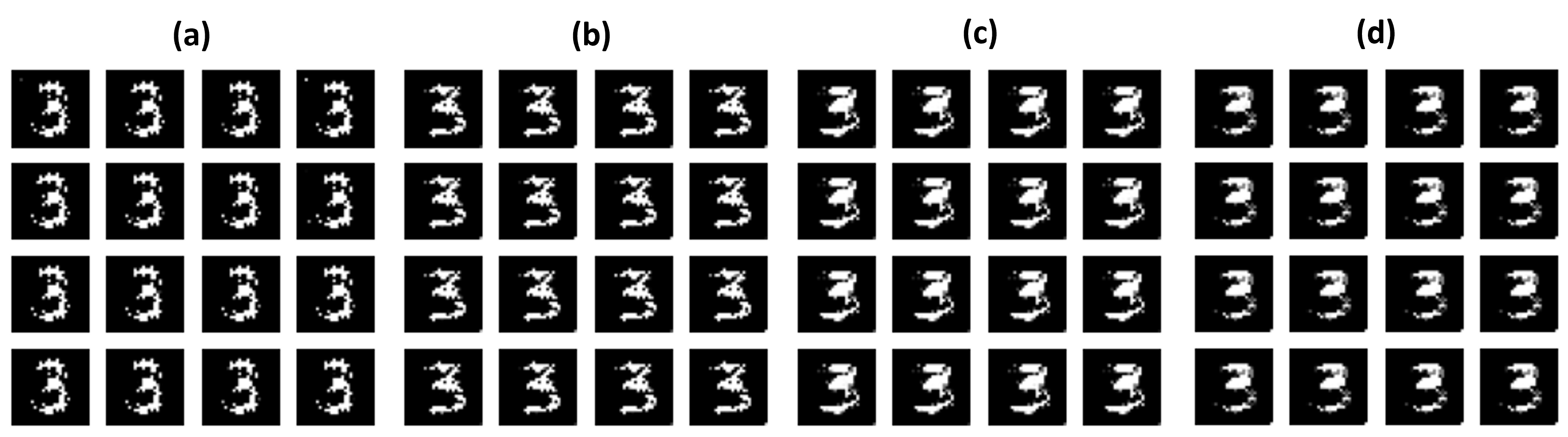}
    \caption{The synthetic images of digit "3" generated by (a) software GAN implementation and the GAN implementation based on the passive RRAM crossbar array utilizing (b) pseudo random normal noise input to the generator network and (c) true random noise input to the generator network without considering the spatial variations (d) true random noise input to the generator network while considering the device-to-device variations.}
\end{figure*}

We use a hardware friendly fixed-amplitude in-situ training methodology, also known as the Manhattan's rule \cite{b29}, which only requires the calculation of sign of the weight gradient ($\Delta W$) for tuning the weights (conductance-states of the RRAM cells encoding weights in the passive crossbar array). Depending on the sign of the gradient and the conductance-state of the RRAM cell, the update rule (shown in Fig. 4) for a positive weight can be given as:

\[
G = 
\begin{cases}
       G_o + \Delta G(G_o,V_{set},t_p),& \text{if } \Delta W>0 \\
       G_o - \Delta G(G_o,V_{reset},t_p),& \text{if } \Delta W<0 \\  
       G_{min},& \text{if } G_o - \Delta G < G_{min} \\
       G_{max},& \text{if } G_o + \Delta G > G_{max}
\end{cases}
\]      

whereas the update rule for a negative weight can be given as:

\[
G = 
\begin{cases}
       G_o - \Delta G(G_o,V_{set},t_p),& \text{if } \Delta W>0 \\
       G_o + \Delta G(G_o,V_{reset},t_p),& \text{if } \Delta W<0 \\  
       G_{min},& \text{if } G_o - \Delta G < G_{min} \\
       G_{max},& \text{if } G_o + \Delta G > G_{max}
\end{cases}
\]      

where, $G_o$ is the conductance-state of the RRAM cell, $\Delta G$ is the absolute change in conductance value, $V_{set} = 0.8V$ and $V_{reset} = - 0.8V$ are the amplitude for set and reset (fixed-amplitude) pulses, respectively, and $t_p$ = 100 ns is the pulse width. For evaluating the change in the conductance value $\Delta G$ upon the application of the fixed-amplitude pulses, we use an experimentally calibrated comprehensive phenomological model for the passive RRAM crossbar array based on the Pt/Al$_2$O$_3$/TiO$_{2-x}$/Ti/Pt stack \cite{b30}. The model not only captures the static characteristics including noise, but also reproduces the experimentally observed dynamic set/reset/conductance-tuning behavior including device-to-device variations and non-linearity for more than 324 RRAMs across $\approx$2 million data points \cite{b30}. The change in the conductance-state $\Delta G$ follows the dynamic equation \cite{b30}:
\begin{align}
    \Delta G = D_{m}(G_0,V_p,t_p) + D_{d2d}(G_0,V_p,t_p)
\end{align}
where $D_{m}$ is the expected noise-free absolute conductance change which depends on the amplitude $V_p$ and duration $t_p$) of the voltage pulse as well as the conductance-state, and $D_{d2d}$ represents the device-to-device variations for different RRAMs on the passive crossbar array.\par

\section{RESULTS AND DISCUSSION}

We perform an extensive analysis of the proposed GAN implementation on passive RRAM crossbar array utilizing the hardware-aware simulation framework developed in section II. We compare the performance of the GAN implementation with different noise-input to the generator in terms of metrics such as accuracy, energy consumption and area. Moreover, to investigate the impact of device-to-device variations on the efficacy of GAN, we analyse the performance of the proposed GAN implementation both in the presence and absence of spatial variations (by appropriately switching the mismatch/variation flag in the compact model).

\begin{figure*}[t]
    \centering
    \includegraphics[width=\textwidth]{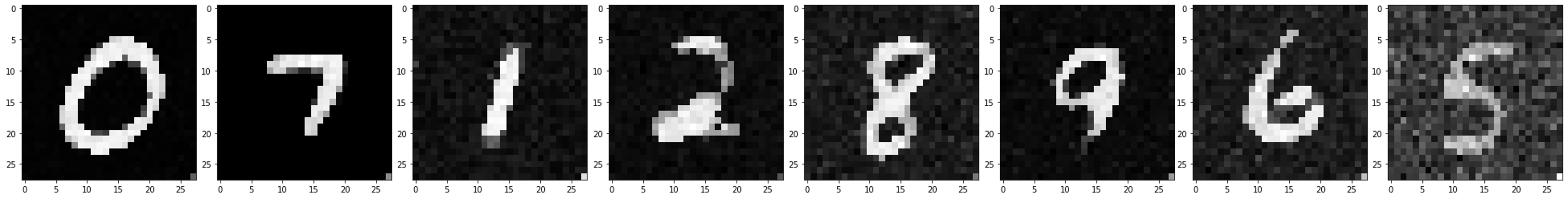}
    \caption{Synthetic images of different digits generated by the proposed GAN implementation on passive RRAM crossbar array exploiting true random noise input to the generator network.}
\end{figure*}

For efficient performance benchmarking, we train (a) software GAN, and the GAN implementations based on passive RRAM crossbar array with (b) psuedo-random noise input, (c) true-random noise input without considering device-to-device variations and (d) true-random noise input considering device-to-device variations for synthesizing digit "3" and divide the training set from MNIST data set into 10 batches (of size 608). The evolution of the conductance-states of the cells in 54 ($64 \times 64$) passive RRAM crossbar arrays during the training of GAN with psuedo-random noise input and true-random noise input in the presence of spatial variations are shown in Fig. 5. It can be observed that different noise inputs to the generator network during training lead to a significantly different conductance-state distribution of the RRAM cells (which represents the weight matrices). This results in synthesized images with different quality and features for different GAN implementations as shown in Fig. 6. Moreover, the representative images generated by the GAN implementation on passive RRAM crossbar array with true random noise input to the generator after training on MNIST dataset are also shown in Fig. 7. As can be observed from Fig. 7, all the classes of synthesized images are not generated with the same accuracy.

\subsection{Accuracy}
Evaluation of the quality of fake images generated by a GAN model is a challenging task. While the method of visual inspection can give a qualitative judgement, several quantitative techniques have also been reported recently \cite{b9}. In this work, we use an image classification-based methodology to estimate the accuracy of the GAN with the aid of features extracted from the generated images. We trained a multi-layer perceptron on the MNIST dataset and used it to extract the features of the fake images and classify them in different class labels. As can be observed from Fig. 8, although the software implementation of GAN shows the highest accuracy after training for 10 batches (1 epoch), the proposed GAN implementation on passive RRAM crossbar array utilizing in-situ training exhibits comparable accuracy when its generator network is fed with true random noise input generated from the RRAM crossbar. Furthermore, the large device-to-device variations in the RRAM switching threshold across the array degrades the accuracy of the GAN implementation.  However, the GAN implementation based on passive RRAM crossbar array with true-random noise input still exhibits a better accuracy as compared to the GAN implementation with pseudo-random noise input even in the presence of hardware imperfections such as device-to-device variations, noise and non-linearity. Moreover, the accuracy of all the GAN implementations converges towards their optimal value after training for large number of batches.

\begin{figure}[h]
    \centering
    \includegraphics[width=0.5\textwidth]{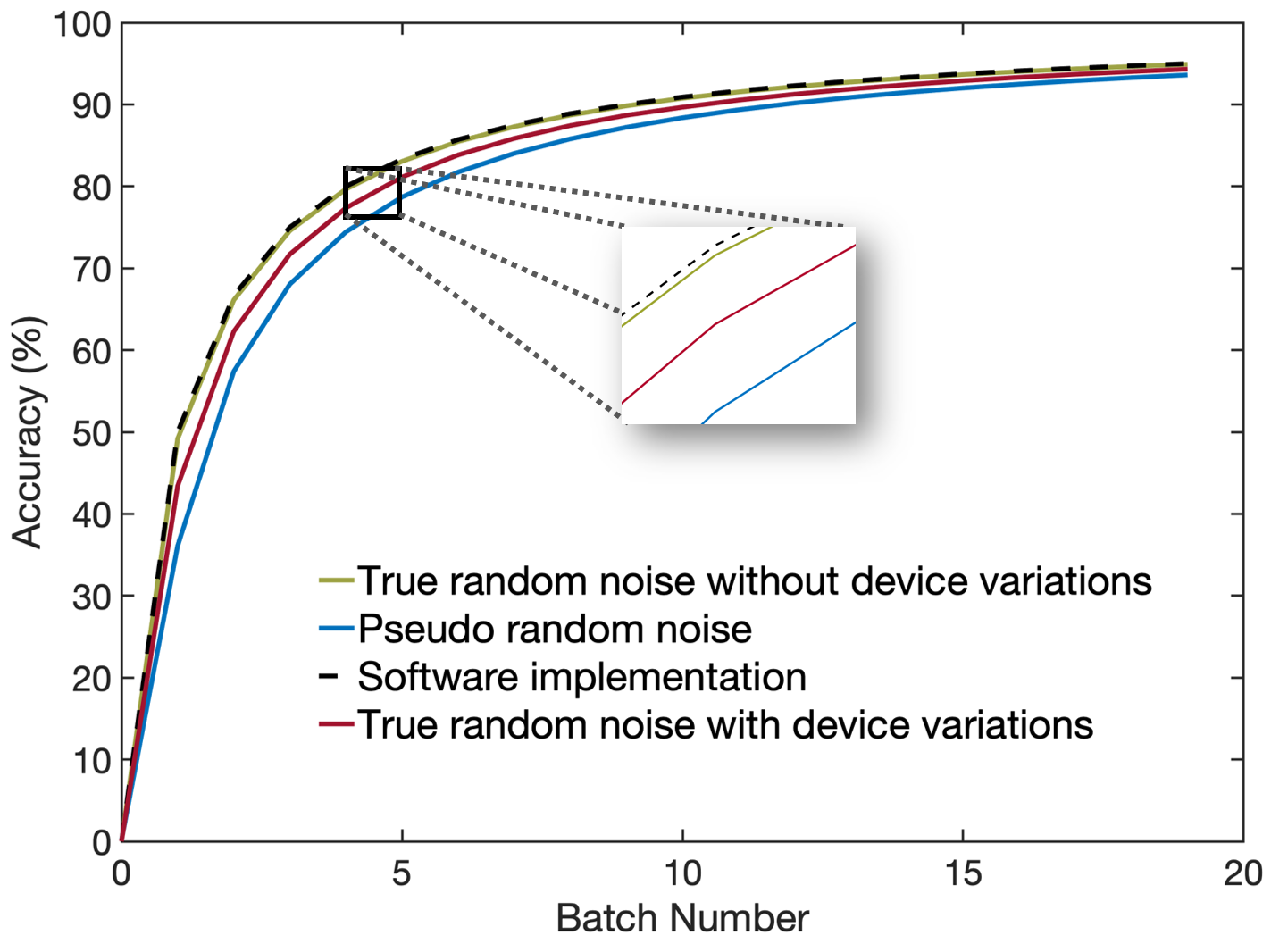}
    \caption{Evolution of the accuracy of different GAN implementations during the training process.}
\end{figure}

\subsection{Energy consumption}

The conventional von-Neumann GAN implementations such as those based on CPUs and GPUs involve frequent data transfer between the memory and processing units significantly increasing their energy consumption and latency. Moreover, simultaneous training of two adversarial networks results in a considerably large number of weight updates during the training process with massive exchange of parameters between the generator and the discriminator. Therefore, training process dominates the energy landscape of the hardware GAN implementations. However, the forward pass and the backward propagation in the proposed GAN implementation based on a passive RRAM crossbar array is inherently non-von-Neumann and significantly reduces the data movement between the processing and storage units increasing its energy-efficiency. Furthermore, the hardware-friendly in-situ training approach utilizing the Manhattan's rule eliminates the need for calculation of the exact values of the weight gradients in the peripheral circuitry further reducing the energy consumption. However, during the in-situ training using Manhattan's rule, set/reset voltage pulses are applied to the RRAM cells in the crossbar array after each batch (weight-update iteration) to change their conductance-state according to the sign of weight gradients which lead to an energy consumption given by: 

$$
    E_i = \begin{cases}
            V_{set}^2 \cdot G_{i-1} \cdot t_p,& \text{if $G_{i} > G_{i-1}$}\\
            V_{reset}^2 \cdot G_{i-1} \cdot t_p,& \text{if $G_{i} < G_{i-1}$}\\
            0,              & \text{otherwise}
        \end{cases}
$$
 
where V$_{set}$ and V$_{reset}$ are the amplitudes of set/reset voltage pulses and t$_p$ is time period of the pulse applied to change the conductance-state from $G_{i-1}$ to $G_i$. Owing to the large number of weight updates, the energy consumed during the conductance-update process dominates the energy landscape of the proposed GAN implementation on passive RRAM crossbar array. 

Utilizing the above equation, the energy consumed during the training process for each batch (weight-update iteration) was calculated for GAN implementations with (a) pseudo-random noise inputs to the generator and true-random noise input to the generator (b) considering spatial variations in the RRAM cells and (c) without considering device-to-device variations as shown in Fig. 9. While the batch-training energy increases with the number of batches for GAN implementation with a pseudo-random noise input to the generator, training the GAN with a true-random noise input to the generator leads to a reduction in the batch-training energy as the training progresses. This can also be inferred from the conductance evolution trends for the two cases as shown in Fig. 5. While the conductance-states change significantly while training the GAN with pseudo-random noise inputs from epoch 10 to epoch 20, the conductance update is rather gradual while training with true-random noise inputs.

The cumulative energy consumption during the training process until the GAN implementations converge to their optimal accuracy are also shown in Fig. 10. The energy consumption is lower when the GAN implementation is trained with true-random noise inputs (48.34 $\mu J$) as compared to the GAN trained with pseudo-random noise input (52.06 $\mu J$). Moreover, device-to-device variations do not lead to a significant change in the energy consumption of the proposed GAN implementation on passive RRAM crossbar array.

\begin{figure}[h]
    \centering
    \includegraphics[width=0.5\textwidth]{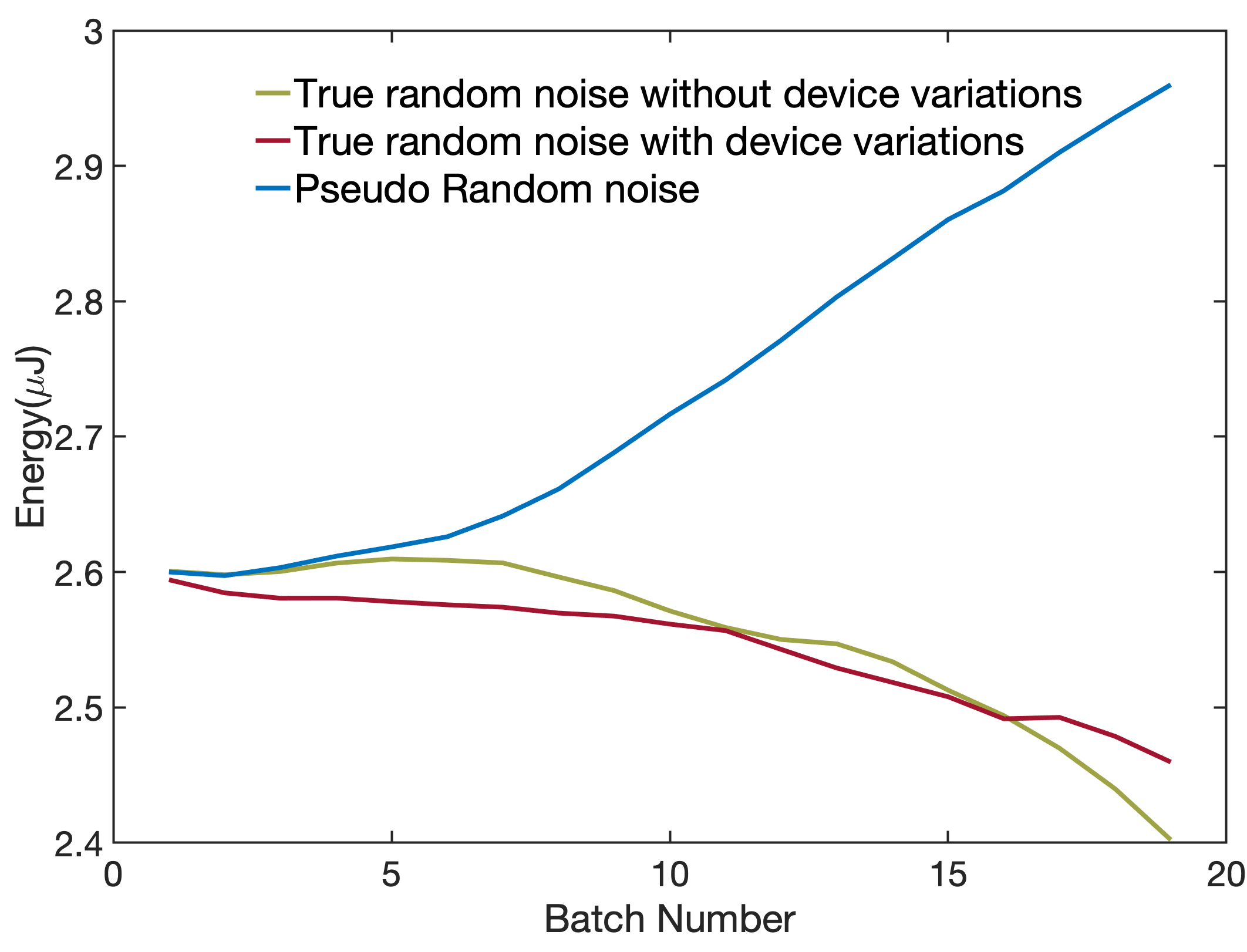}
    \caption{Energy dissipation in the passive RRAM crossbar array in different batches while training the different GAN implementations.}
\end{figure}

\begin{figure}[h]
    \centering
    \includegraphics[width=0.5\textwidth]{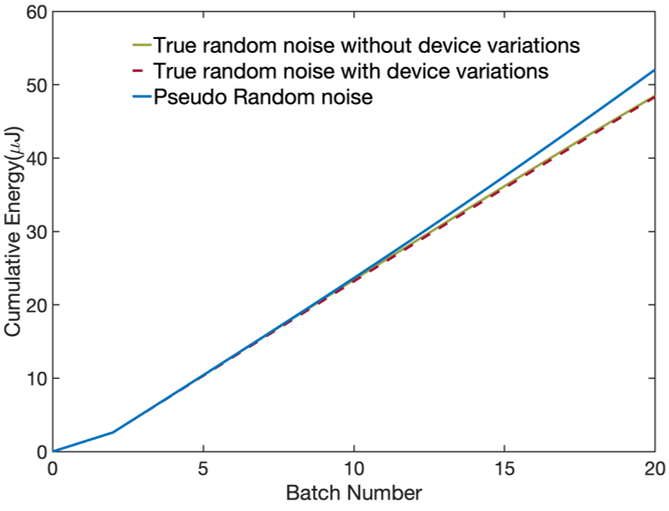}
    \caption{Cumulative energy dissipated in the passice RRAM crossbar array while training the different GAN implementations.}
\end{figure}

\subsection{Area}

For practical applications such as image super-resolution, etc., GANs need to generate high resolution images which requires large weight matrices between the different layers of the adversarial networks. Therefore, highly scalable and compact realization of synaptic elements (representing weights) is crucial for area-efficient hardware GAN implementations. Since a single RRAM cell in a passive RRAM crossbar array occupies an area of 0.36 $\mu m^2$ (0.6 $\mu m$ $\times$ 0.6 $\mu m$), a 64$\times$64 crossbar has a footprint of 1474.56 $\mu m^2$. The total area occupied by the proposed GAN implementation with 54 such crossbars is 0.079 $mm^2$ which is extremely low as compared to the area occupied by the active 1T-1R array of similar size \cite{b10}. Moreover, the area-efficiency of the proposed implementation can be further enhanced by utilizing 3D-integration of several layers of RRAM cells \cite{b31}.

\section{Conclusions}
In this work, we have proposed a highly scalable, compact and energy-efficient GAN accelerator which performs the key operations such as forward pass and backward propagation, training and noise generation in-situ on a passive RRAM crossbar array. Unlike the prior GAN implementations, we have also evaluated the impact of the noise input used for training the generator network on the performance of GAN. Our extensive investigation utilizing an experimentally calibrated phenomenological model for passive RRAM crossbar array reveals that training GAN with a true-random noise input leads to a significant reduction in the training energy without degrading the accuracy. Our results may encourage experimental demonstration of GAN accelerators on passive RRAM crossbar arrays.


\begin{thebibliography}{00}
\bibitem{b1} I. Goodfellow, J. Pouget-Abadie, M. Mirza, B. Xu, D. Warde-Farley, S. Ozair, A. Courville, and Y. Bengio, "Generative adversarial nets," \emph{Advances in neural information processing systems}, pp. 27, 2014. doi:10.5555/2969033.2969125.
\bibitem{b2} A. Creswell, T. White, V. Dumoulin, K. Arulkumaran, B. Sengupta, and A. A. Bharath, "Generative adversarial networks: An overview. \emph{IEEE Signal Processing Magazine}, vol. 35, no. 1, pp.53-65, 2018. doi:10.1109/MSP.2017.2765202.
\bibitem{b3} N. Shrivastava, M. A. Hanif, S. Mittal, S. R. Sarangi, and M. Shafique, "A survey of hardware architectures for generative adversarial networks. \emph{Journal of Systems Architecture}, vol. 118, p.102227, 2021. doi:10.1016/j.sysarc.2021.102227.
\bibitem{b4} F. Chen, L. Song and Y. Chen, "ReGAN: A pipelined ReRAM-based accelerator for generative adversarial networks," 2018 \emph{23rd Asia and South Pacific Design Automation Conference (ASP-DAC)}, 2018, pp. 178-183, doi: 10.1109/ASPDAC.2018.8297302.
\bibitem{b5} Z. Fan, Z. Li, B. Li, Y. Chen and H. Li, "RED: A ReRAM-based Deconvolution Accelerator," \emph{2019 Design, Automation and Test in Europe Conference and Exhibition (DATE)}, 2019, pp. 1763-1768, doi: 10.23919/DATE.2019.8715103.
\bibitem{b6}F. Chen, L. Song, H. Li and Y. Chen, "ZARA: A Novel Zero-free Dataflow Accelerator for Generative Adversarial Networks in 3D ReRAM," 2019 56th ACM/IEEE Design Automation Conference (DAC), 2019, pp. 1-6.
\bibitem{b7} A. S. Rakin, S. Angizi, Z. He and D. Fan, "PIM-TGAN: A Processing-in-Memory Accelerator for Ternary Generative Adversarial Networks," \emph{2018 IEEE 36th International Conference on Computer Design (ICCD)}, 2018, pp. 266-273, doi: 10.1109/ICCD.2018.00048.
\bibitem{b8} O. Krestinskaya, B. Choubey, and A. P. James, "Memristive GAN in analog." \emph{Scientific reports}, vol. 10, no. 1, pp.1-14, 2020. doi:10.1038/s41598-020-62676-7.
\bibitem{b9}Y. Lin et al., "Demonstration of Generative Adversarial Network by Intrinsic Random Noises of Analog RRAM Devices," 2018 IEEE International Electron Devices Meeting (IEDM), 2018, pp. 3.4.1-3.4.4, doi: 10.1109/IEDM.2018.8614483.

\bibitem{b10} H. Wu et al., "Device and circuit optimization of RRAM for neuromorphic computing," 2017 IEEE International Electron Devices Meeting (IEDM), 2017, pp. 11.5.1-11.5.4, doi: 10.1109/IEDM.2017.8268372.
\bibitem{b11} P. Yao, H. Wu, B. Gao, J. Tang, Q. Zhang, W. Zhang, J.J. Yang and H. Qian, “Fully hardware-implemented memristor convolutional neural network,” \emph{Nature}, vol. 577, pp.641-646, 2020. doi:10.1038/s41586-020-1942-4.
\bibitem{b12} M. Hu, C.E. Graves, C. Li, Y. Li, N. Ge, E. Montgomery, N. Davila, H. Jiang, R.S. Williams, J.J. Yang and Q. Xia, “Memristor‐based analog computation and neural network classification with a dot product engine,” \emph{Advanced Materials}, vol. 30, 2018. doi:10.1002/adma.201705914.
\bibitem{b13} S. Ambrogio, S. Balatti, V. Milo, R. Carboni, Z.Q. Wang, A. Calderoni, N. Ramaswamy and D. Ielmini, “Neuromorphic learning and recognition with one-transistor-one-resistor synapses and bistable metal oxide RRAM,” \emph{IEEE Transactions on Electron Devices}, vol. 63, no. 4, pp.1508-1515, 2016. doi: 10.1109/TED.2016.2526647.
\bibitem{b14} F. Cai, J.M. Correll, S.H. Lee, Y. Lim, V. Bothra,  Z. Zhang, M.P. Flynn and W.D. Lu, “A fully integrated reprogrammable memristor–CMOS system for efficient multiply–accumulate operations,” \emph{Nature Electronics} vol. 2, pp.290-299, 2019. doi:10.1038/s41928-019-0270-x.

\bibitem{b15} H. Kim, H. Nili, M. Mahmoodi and D. Strukov, “4K-memristor analog-grade passive crossbar circuit,”\emph{Nat. Comm.}, vol. 12, no. 1, pp.1-11, 2021. doi:10.1038/s41467-021-25455-0.
\bibitem{b16} M. Prezioso, et al. Training and operation of an integrated neuromorphic network based on metal-oxide memristors. Nature 521, 61–64 (2015). https://doi.org/10.1038/nature14441.
\bibitem{b17} F. Alibart, E. Zamanidoost, and D. Strukov, “Pattern classification by memristive crossbar circuits using ex situ and in situ training,” \emph{Nat. comm.}, vol. 4, no. 1, pp.1-7, 2013. doi:10.1038/ncomms3072.
\bibitem{b18} F. M. Bayat, M. Prezioso, B. Chakrabarti, H. Nili, I. Kataeva and D. Strukov, “Implementation of multilayer perceptron network with highly uniform passive memristive crossbar circuits,” \emph{Nat. comm.}, vol. 9, no. 1, pp.1-7, 2018. doi:10.1038/s41467-018-04482-4.
\bibitem{b19} P. M. Sheridan, F. Cai, C. Du, W. Ma, Z. Zhang and W. D. Lu, “Sparse coding with memristor networks”, \emph{ Nature nanotechnology}, 12(8), pp.784-789, 2017. doi:10.1038/nnano.2017.83.
\bibitem{b20} H. Yeon, P. Lin, C. Choi, S. H. Tan, Y. Park, D. Lee, J. Lee, F. Xu, B. Gao, H. Wu, H. Qian, Y. Nie, S. Kim and J. Kim, “Alloying conducting channels for reliable neuromorphic computing”, \emph{  Nat. Nanotechnol} 15, 574–579 (2020). doi:10.1038/s41565-020-0694-5.
\bibitem{b21} H. Nikam, S. Satyam and S. Sahay, "Long Short-Term Memory Implementation Exploiting Passive RRAM Crossbar Array," in \emph{IEEE Transactions on Electron Devices}, doi: 10.1109/TED.2021.3133197.
\bibitem{b22} M. Hu, J. P. Strachan, Z. Li, E. M. Grafals, N. Davila, C. Graves, S. Lam, N. Ge, J. J. Yang and R. S. Williams, ‘‘Dot-product engine for Neuromorphic computing: Programming 1T1M crossbar to accelerate matrix-vector multiplication,’’ in \emph{Proc.
53rd ACM/IEEE Design Automat. Conf. (DAC)}, pp. 1–6, 2016. doi:10.1145/2897937.2898010.
\bibitem{b23} M. J. Marinella et al., "Multiscale Co-Design Analysis of Energy, Latency, Area, and Accuracy of a ReRAM Analog Neural Training Accelerator," \emph{in IEEE Journal on Emerging and Selected Topics in Circuits and Systems}, vol. 8, no. 1, pp. 86-101, 2018. doi: 10.1109/JETCAS.2018.2796379. 
\bibitem{b24} S. Sahay, M. Bavandpour, M. R. Mahmoodi and D. Strukov, “A 2T-1R cell array with high dynamic range for mismatch-robust and efficient neurocomputing,” in \emph{proc. IEEE Int. Memory Workshop (IMW)}, pp. 1-4, 2020. doi:10.1109/IMW48823.2020.9108142.
\bibitem{b25} M. Bavandpour, S. Sahay, M. R. Mahmoodi and D. Strukov, "Efficient mixed-signal neurocomputing via successive integration and division," \emph{IEEE Trans. VLSI systems}, vol. 28, no. 3, pp. 823-827, 2020. doi:10.1109/TVLSI.2019.2946516.

\bibitem{b26} H. Nili, G. C. Adam, B. Hoskins, M. Prezioso, J. Kim, M. R. Mahmoodi, F. M. Bayat, O. Kavehei, and D. B. Strukov, "Hardware-intrinsic security primitives enabled by analogue state and nonlinear conductance variations in integrated memristors," \emph{Nature} Electronics, vol. 1, no. 3, pp.197-202, 2018. doi: 10.1038/s41928-018-0039-7.
\bibitem{b27} S. Sahay, A. Kumar, V. Parmar, and M. Suri, "OxRAM RNG circuits exploiting multiple undesirable nanoscale phenomena," \emph{IEEE Transactions on Nanotechnology}, vol. 16, no. 4, pp.560-566, 2017. doi: 10.1109/TNANO.2016.2647623.
\bibitem{b28} S. Sahay, and M. Suri, "Recent trends in hardware security exploiting hybrid CMOS-resistive memory circuits," \emph{Semiconductor Science and Technology}, vol. 32, no. 12, p.123001, 2017. doi: 10.1088/1361-6641/aa8f07.
\bibitem{b29} I. Kataeva, F. Merrikh-Bayat, E. Zamanidoost and D. Strukov, “Efficient training algorithms for neural networks based on memristive crossbar circuits,” in \emph{proc. IEEE Int. Joint Conf. Neural Networks (IJCNN)}, pp. 1-8, 2015. doi: 10.1109/IJCNN.2015.7280785.

\bibitem{b30} H. Nili, A. F. Vincent, M. Prezesio, M. R. Mahmoodi, I. Kataeva and D. B. Strukov, “Comprehensive compact phenomenological modeling of integrated metal-oxide memristors,” \emph{IEEE Trans. Nanotechnology}, vol. 19, pp. 344-349, 2020. doi: 10.1109/TNANO.2020.2982128.
\bibitem{b31} G. C. Adam, B. D. Hoskins, M. Prezioso, F. Merrikh-Bayat, B. Chakrabarti and D. B. Strukov, “3-D memristor crossbars for analog and neuromorphic computing applications,” \emph{IEEE Trans. Electron Devices}, vol. 64, no. 1, pp.312-318, 2016. doi:10.1109/TED.2016.2630925.


\end{thebibliography}
\end{document}